\documentclass[12pt]{article}
\usepackage{epsfig}

\usepackage{amsmath}
\usepackage{amssymb}
\usepackage{euscript}

\begin{document}

\begin{titlepage}
 
\begin{flushright}
\bf IFJ-PAN-Report-1981/PH
\end{flushright}
\vspace{2mm}

\begin{center}      
{\LARGE\bf Luminosity Measurement Method for  LHC:}    \\ 
\vspace{15mm}
{\Large\bf The theoretical precision and the experimental challenges$^{\star}$}
\vspace{2mm}
\end{center}

\vspace{5mm}

\begin{center}
{\large\bf  M.~W. Krasny$^{a,b}$, J. Chwastowski$^{b}$, K. S{\l}owikowski$^{c}$}

\vspace{4mm}
{\em $^a$LPNHE, Pierre and Marie Curie University, Tour 33, RdC, \\ 
                4, pl. Jussieu 75005 Paris, France.}\\
 \vspace{2mm}
{\em $^b$Institute of Nuclear Physics HNINP-PAS,\\
  ul.\ Radzikowskiego 152, 31-342 Cracow, Poland.} \\
 \vspace{2mm}
{\em $^c$Faculty of Physics and Applied Computer Science, \\
 AGH-University of Sciences and Technology, \\
 Al. \ Mickiewicza 30, 30-059 Cracow, Poland.}

\end{center}

\vspace{5mm}
\begin{abstract}

This is the first of the series of papers which present a precision method 
of the day-by-day  monitoring of the absolute LHC luminosity. This method is 
based on the measurement of the rate of coplanar lepton pairs
produced in peripheral collisions of the beams' particles.
In the present paper we evaluate the modeling precision of the 
lepton pair production processes in proton-proton collisions, 
optimize the measurement region
to achieve better than 1\%  accuracy of the predicted rates, and discuss
the experimental challenges to filter out the luminosity monitoring lepton 
pairs at LHC.
\end{abstract}

\vspace{3mm}
\begin{center}
\end{center}

\vspace{2mm}
\footnoterule
\noindent
{\footnotesize
$^{\star}$The work is supported by the program of cooperation 
between the IN2P3 and Polish Laboratories No.\ 05-117 and by the POLONIUM  
program No.\ 11580PE.
}

\end{titlepage}

\section{Introduction}
\label{sec:Introduction}

The Large Hadron Collider (LHC) experimental program is mainly focused on the 
study of collisions of point-like constituents of the LHC beam particles. If 
their collision rates could be unambiguously expressed in terms of the 
scale-dependent, universal  partonic luminosities and process-specific cross 
sections then the measurement of the parent beam luminosity would be of 
residual importance for such a program. However, the unfolding of partonic 
luminosities involves modeling of the confinement effects of virtual bunches of
partons and the effects of strongly interacting medium which neither can  be 
controlled by the perturbative QCD nor can be proved to be collision-energy 
independent. Because of these facts the only precise, model independent method 
of the absolute normalisation of the rates of partonic collisions must remain 
to be based on the measured luminosities of the parent beam particles: protons 
and ions. 

Numerous methods of the LHC luminosity measurement have been proposed -  
see \cite{MARTIN} and references quoted therein for a comprehensible review.
In general, all the proposed luminosity measurement methods are based on:
\begin{enumerate}
\item 
the measured properties of the colliding beams: the beam
currents and profiles. These methods provide the luminosity delivered to the 
interaction point and need to be corrected for the dead-time of the trigger 
and the data acquisition systems to determine the recorded luminosity. 
The expected accuracy of such methods is of the order of 10\%  \cite{LHC},    
\item
a simultaneous measurement of the rate of two processes  
inter-linked via the optical theorem. These methods can be applied only in the 
dedicated collider runs with high $\beta^{*}$ beam optics.  The luminosity 
measurement must be subsequently extrapolated to the standard low $\beta^{*}$ 
ones. The expected measurement accuracy in the high $\beta^{*}$ mode is 2-3\% 
\cite{ATLAS}. The precision of the extrapolation will depend upon the machine 
operating conditions and, at present, cannot be reliably estimated,
\item
the measurement of the rates of the electromagnetic processes. The precision 
of these methods is in the range of 2-5\%. It reflects the process dependent
resolving power of the internal charge structure of the beam particles.
Several methods minimizing the impact of the internal charge structure of the
colliding particles have been elaborated by the HERA collider experiments 
\cite{H1RADIATIVE}, and could be used at LHC.
\item
a dedicated machine runs with hybrid beams \cite{ELECTRN_BEAM}. In these runs 
a parasitic electron beam carried to the interaction point by the ions could 
be used for the measurement of the absolute machine luminosity. A precision of 
1-2\% similar to that of the HERA collider \cite{H1RADIATIVE,ZEUS} can be 
achieved.
\end{enumerate}

The discussion of the methods belonging to each of these four categories is
beyond the scope of this paper. This and the subsequent papers document an 
attempt to select the method which, in our view, has the largest potential of 
providing the highest precision at the LHC, and to develop a proposal of the 
complete {\bf luminosity measurement procedure} which includes: high-precision 
modeling of the signal and background processes, the proposals of the dedicated
detector and the trigger, construction of the on- and off-line data selection 
and the data-monitoring  algorithms, and last, but not least, elaboration of 
the concrete luminosity counting methods.

Our primary goal is to achieve a $\leq$1\% statistical, theoretical and 
systematic precision of the  absolute normalisation of the  collected event 
samples on a daily basis, and to provide handy tools for measuring the cross 
sections for any user selected event sample, regardless of their sampling 
frequency and the luminosity-block-based data segmentation. Moreover, our aim 
is to provide dedicated tools for a high precision on-line monitoring of the 
bunch-by-bunch relative luminosity at the time granularity of a couple of 
seconds for an efficient machine operation feedback. Last but not the least,
our goal is to develop a method  which is universal for any beam species 
colliding at the LHC, protons and ions. This last point is particularly 
important for a high precision scrutinizing of the polarisation asymmetry in 
the collisions of effective beams of polarised W-bosons with hadronic matter 
discussed in \cite{WBEAMS}.

It is obvious that the highest precision of the luminosity measurement for the 
colliding beams of strongly interacting composite particles could be achieved
by counting the rate of their highly peripheral collisions. In such collisions 
the internal structure of the beam particles and their strong interactions may 
be neglected. Thus, their rate can be precisely calculated using exclusively 
the QED perturbative methods. It can be expressed solely in terms of the static
properties of the beam particles: their masses, electric charges and magnetic 
moments - the parameters which are known to a very high precision \cite{PDG}.

The simplest ``luminometric'' process which could satisfy the above condition 
is the small angle, elastic scattering of the beam particles. However, for the 
high luminosity (small $\beta ^{*}$) operation mode of the ring-ring colliders 
the highly-peripheral, elastic collisions cannot be efficiently selected. 
A remedy, elaborated and employed successfully at HERA, was to select  the
process of quasi-elastic {\it radiative scattering} 
$$ep \rightarrow e + \gamma + p$$
and use the radiative photon as the collision peripherality tagger.
Only a fraction of radiative photons can play such a role. In the HERA collider
case the photon  must  be emitted co-linearly to the incoming electron (the 
bremsstrahlung process  \cite{SUSZYCKI}) or  its  transverse momentum must 
balance the electron transverse momentum (the Compton process  \cite{COURAU}). 

An equivalent  process for hadronic colliders is the quasi-elastic lepton pair 
production process 
$$pp \rightarrow l^+l^- + pp$$
in which the lepton pair is emitted co-linearly to the beams' collision axis.
The luminosity measurement method presented in this and subsequent papers
employs such pairs for the tagging of peripheral processes at LHC in the same 
manner as the radiative photons were employed at HERA. The above analogy 
enables us to make use of our experience gained at HERA \cite{H1RADIATIVE,ZEUS}
while developing a  proposal of the luminosity measurement procedure  for LHC.
Apart from the differences in the detection schemes of the photons and of the 
lepton pairs the measurement procedure will essentially remain the same as that
developed and tested at HERA.
 
This paper is the first of the series of papers presenting such a proposal. It 
is organized as follows. In section 2 we describe the experimental signatures
and the modeling of the lepton pair production processes. We concentrate on the
proton--proton case leaving the discussion of the ion--proton and ion--ion 
collisions to a dedicated paper \cite{paper4}. In section 3 we discuss the 
optimisation of the measurement region to achieve the highest statistical and
theoretical precision of the predicted event rates. In section 4  we present 
the experimental challenges which the  measurement of the rate of lepton pair 
production in the selected phase-space region must face. 

\section{Lepton pair production at LHC }
\label{sec:Luminometer}

\subsection{Phase-space region selection} 
\label{sec:Signatures }
 
Budnev, Ginzburg, Meledin and Serbo \cite{Budneev} were first to propose the 
lepton pair production process  
for a precise determination of the proton--proton collision luminosity at  
high energy colliders.

This process has  already been studied as a candidate process to measure 
luminosity at LHC. Production of electron--positron pairs in the central 
pseudorapidity region was studied by Telnov \cite{Telnov}. These studies were 
abandoned because of a lack of a viable triggering scheme for such pairs. 

The detection of forward-produced $e^+e^-$ pairs with a very small 
invariant mass, $M_{ll} \leq 0.01 GeV$, at the distance of 19 m from the LHC 
interaction point,  was proposed  in \cite{Piotrzkowski} and \cite{ATLAS_TDR}.
The measurement in this phase-space region is difficult because of the limited 
experimental means of controlling the beam--beam and beam--wall (beam--gas) 
collision background. In addition,  such a measurement requires not only a 
precise mapping of the magnetic field and the dead material over the distance 
of 19 meters, but also a precise knowledge of the beam angular divergence. Last
but not the least, the event rates in this kinematic region are sensitive to 
bunch dynamics, in particular to the space-charge effects of the LHC bunches. 
This method has also been abandoned.

Production of muon pairs in the central pseudorapidity region, proposed first 
by A. Courau \cite{Courau}, has been studied  in \cite{Shamov}. The authors of 
\cite{Shamov} argued that this process could be used to determine the LHC 
luminosity with $\sim$2\%  systematic precision. To achieve such a precision 
the machine and the beam-collision background must, however, be small enough 
to allow for an efficient triggering of muons down to the transverse momenta  
of $p_t^{\mu}~ = ~6$  GeV/c. The most important limitation of this method is, 
however, its statistical precision. The cross section for production of the 
muon pair of high invariant mass is small and 1\% statistical precision of the 
luminosity measurement can be achieved only on the year-by-year, rather than 
on the day-by-day,  basis  at the machine luminosity of 
${\cal L} = 10^{33} cm^{-2}s^{-1}$.
\begin{figure}
\begin{center}
\setlength{\unitlength}{1mm}
\begin{picture}(170,80)
\epsfig{file=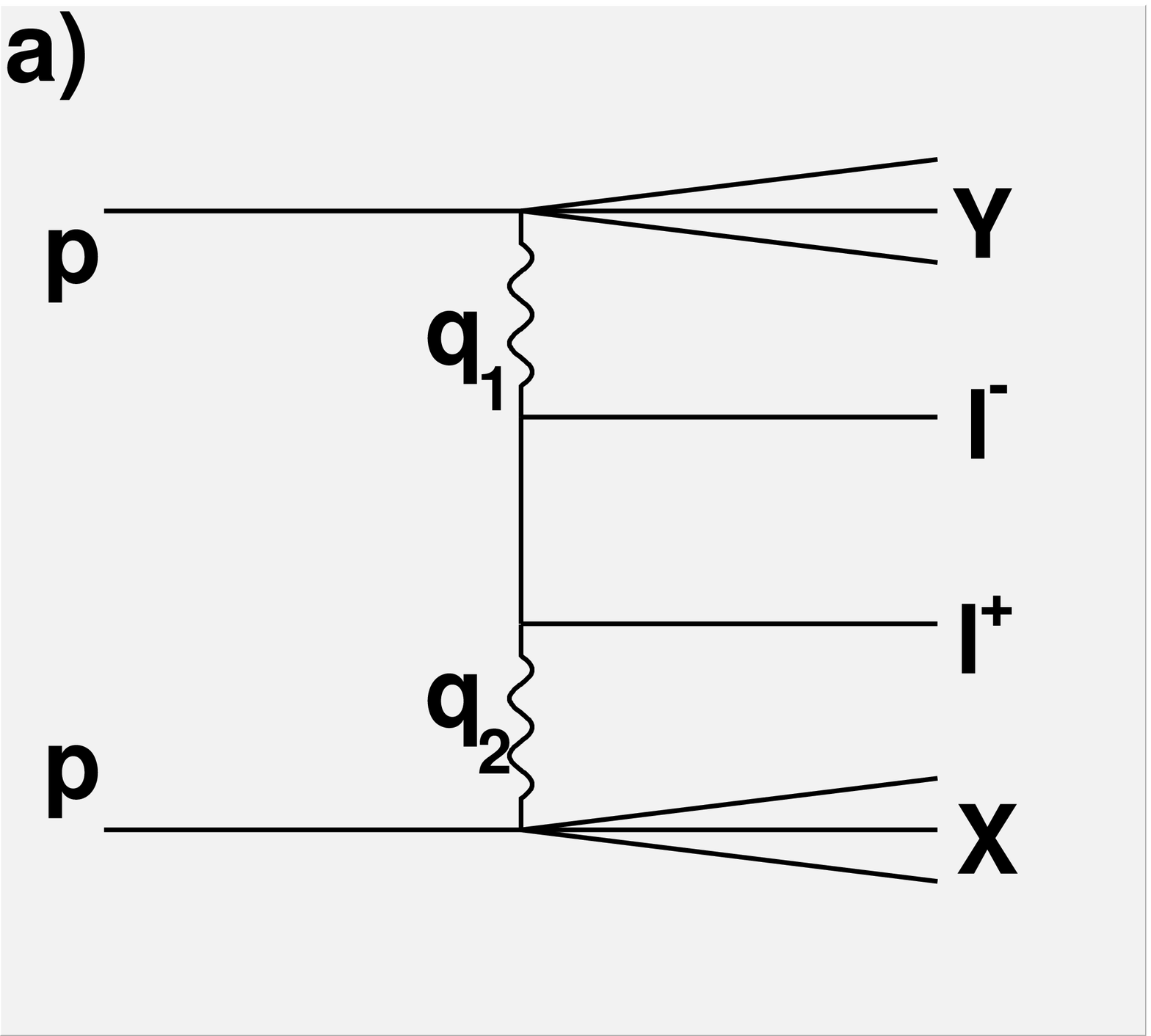,width=70mm,height=70mm} \hspace{7mm}
\epsfig{file=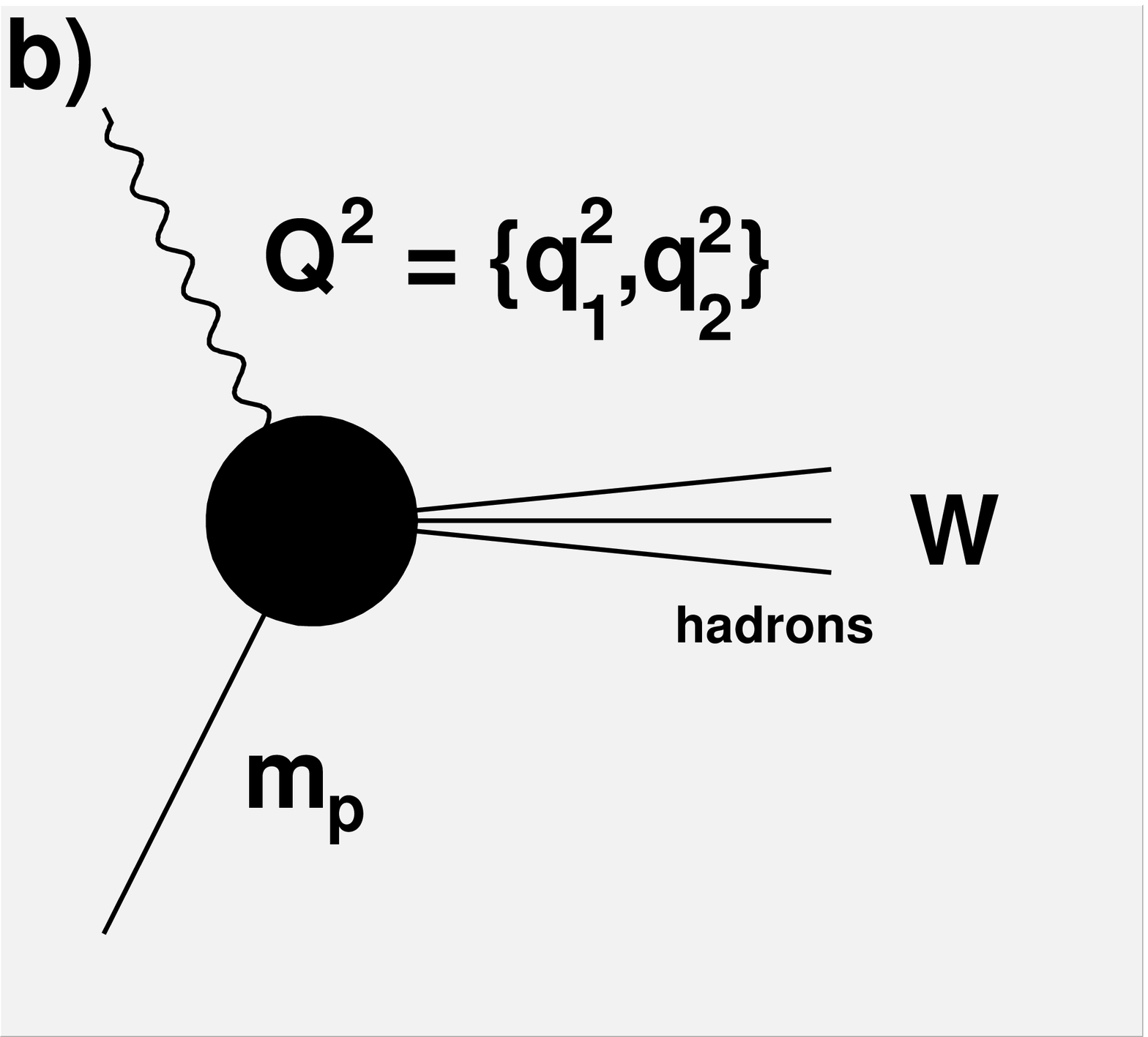,width=70mm,height=70mm}
\end{picture}
\end{center}
\caption{\sf (a) Electromagnetic production of lepton pairs in 
proton--proton collision. X and Y represent either the remnants of the protons 
produced in inelastic collisions  or the outgoing protons 
for the elastic ones, (b) the vertex of the electromagnetic probe of the 
proton structure.}
\label{diagram}
\end{figure}

For the method presented here we choose the phase-space region of the invariant
mass of the lepton pair $M_{ll} \geq 0.4$ GeV. We focus our attention on those 
of the beam collisions in which the lepton pair is produced exclusively in the 
pseudorapidity region $-2.7 < \eta < 2.7$. In the selected region of
centrally-produced and a relatively large invariant mass\footnote{The invariant
mass of the lepton pair is 1000 times higher than the electron mass.} pairs 
each of the two leptons is produced at sufficiently large transverse momentum 
to be independent of the coherent bunch-space-charge effects, and to neglect 
the proton beams' crossing angle and divergence. Moreover, as far as the  
electron-positron pairs are concerned, the Dalitz pair background is 
suppressed. Since the strong interactions of colliding  protons are attenuated 
at  the distance scales larger than the typical distance scales of strong 
interactions: $1/m_{\pi}, 1/\Lambda$,  the highly peripheral lepton pairs  
($p_t^{pair} \ll 1/m_{\pi}, 1/\Lambda$) are  produced essentially in 
electromagnetic processes.

The residual contribution of strong interaction processes is further reduced by
demanding the presence of rapidity gaps between the lepton pair and the 
remnants of the incident protons. Such a requirement assures not only the 
direct electromagnetic coupling of leptons to protons but removes 
simultaneously the processes in which the lepton pair comes from the decays of 
produced hadrons. Finally, the choice of centrally produced pairs removes the 
contribution of radiative lepton pair production by the beam particles and by 
their low mass excitations originating from the  Pomeron mediated collisions.

For the lepton pairs satisfying the above requirements the photon--photon 
fusion, shown in Fig. \ref{diagram}a,  is the dominant Born-level process. The 
higher order corrections can be classified as the strong interaction and the 
electromagnetic radiative corrections. The former have been discussed in 
\cite{MARTIN} and were found to be negligible for the the low $p_t$ lepton 
pairs. The latter are small \cite{Budneev} and can be controlled at the LHC 
energies with precision better than 0.1\% \cite{Jadach}.

\subsection{Modeling of the lepton pair production  processes} 
\label{sec:Modeling }

\subsubsection{The LPAIR generator}

The LPAIR event generator \cite{Baranow} was used to simulate the lepton pair 
production events. It is based on a computation by J.A.M. Vermaseren 
\cite{Vermaseren}. This is an extension of an  earlier work on the lepton pair 
production in $e^+e^-$ collisions by P. Kessler \cite{Kessler} and includes, as
an option, the formalism for collisions of composite beam particles with  an 
arbitrary charge structure. 

In the LPAIR generator the lepton pair production cross section is calculated 
as a convolution of the fluxes of virtual photons and the cross section for 
virtual photon collisions:
\begin{eqnarray} 
 \sigma (p_{1}, p_{2}) = \sigma_{\gamma ^{*} \gamma ^{*} \rightarrow l_1 l_2} 
(p_{1}, p_{2},q_1,q_2) \otimes   flux _{\gamma ^{*}} (q_1)  \otimes 
flux_{\gamma ^{*}}(q_2), 
\end{eqnarray}
where $p_{1}$ and $p_{2}$ represent the four-momenta of the outgoing, on-shell
leptons,  and $q_1$ and $q_2$ the four-momenta of virtual photons. In our 
analysis we use both the LPAIR matrix element for the $\gamma^{*}\gamma^{*}$ 
collisions and the LPAIR algorithm for the convolution of the corresponding 
cross section with virtual photon fluxes. However, we have upgraded the photon 
flux modeling. Such an upgrade was indispensable since, as we shall discuss in 
the following sections, the calculation precision of the event rates was mainly
limited by approximations used in the photon flux modeling rather than by the 
accuracy of the theoretical calculations and the precision of numerical 
algorithms. It allowed us to study the effects which were neglected in earlier 
analyses. 

\subsubsection{Photon fluxes}

\begin{figure}
\begin{center}
\epsfig{file=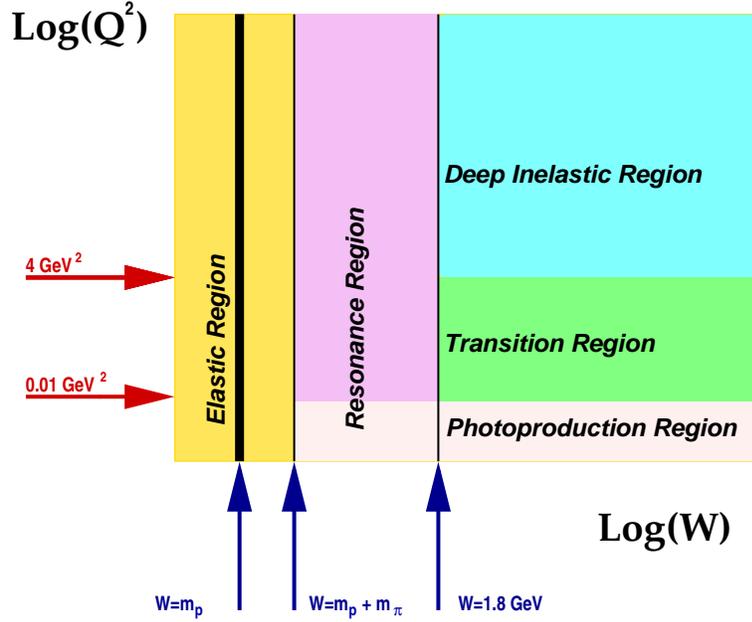,width=10cm}
\end{center}
\caption{\sf The kinematic regions used in the modeling of the photon fluxes.}
\label{Regions}
\end{figure}

The coupling of a virtual photon to a proton, shown in Fig. \ref{diagram}b, is
specified in terms of the proton mass, $m_p$, the produced hadronic system 
mass, $W= \sqrt{(\Sigma p_i)^2 }$, where $p_i$ are the four-momenta of all 
outgoing hadrons,  and the four-momentum transfer squared, $Q^2$. These 
Lorenz-invariant variables provide a complete description of the photon--proton
coupling for inclusive  processes in  which the corresponding cross section is 
integrated over all possible hadronic configurations and summed over the 
polarisation states of virtual photons. In order to model the photon--proton 
coupling we split the $(Q^2,W)$ plane into five regions: the elastic region in 
which $W=m_p$, the resonance region defined by the condition 
$m_p+m_{\pi} \leq W \leq 1.8$  GeV, the photoproduction region satisfying the 
condition $Q^2 \leq 0.01$ GeV$^2$, the deep inelastic region in which  
$Q^2 \geq 4$ GeV$^2$, and the transition region spread over the remaining part 
of the $(Q^2,W)$ plane. These regions are shown schematically in Fig. 
\ref{Regions}. 

In the elastic region the photon fluxes are specified in terms of the form 
factors $G_E(Q^2)$ and $G_M(Q^2)$ which have been re-measured at SLAC 
\cite{E140}. In the resonance region we use the parameterised form 
\cite{DESY} of the DESY low energy data for the photo-absorption cross section 
of longitudinally and transversely polarised photons. In the photoproduction 
and the transition regions we use  the parameterisation of \cite{Levy}. This 
parameterisation interpolates the data between the photoproduction and the deep
inelastic regions and assures a smooth transition to the resonance region. In 
the deep inelastic region we use the MRS \cite{MRS} parameterisation of the 
partonic distribution functions.

\subsection{Modeling precision}

The precision of the LPAIR  predictions for the lepton pair rates depends on
the matrix element calculation accuracy, the size of the higher order 
corrections to the process shown in Fig.\ref{diagram}a, and the modeling 
accuracy  of the virtual photon fluxes.
For the luminosity measurement method discussed in this paper the accuracy of 
photon flux modeling determines entirely the precision of the predicted lepton 
pair rates. If ordered according to the increasing modeling precision the 
lepton pair production processes can be grouped into four classes:
\begin{enumerate} 
\item 
the inelastic processes sensitive to the momentum distributions of quarks in 
the proton (a $\sim$5 \% precision),
\item 
the inelastic processes sensitive to the collective excitations of quarks in 
the proton (a $\sim$4\% precision),
\item
the elastic processes sensitive to the proton magnetic and electric form 
factors (a $\sim$2\% precision),
\item
the elastic processes in which proton behaves as point-like particle; sensitive
only to the proton mass, charge and anomalous magnetic moment which are known 
to a  very high accuracy \cite{PDG}.
\end{enumerate}

In general each of  the above classes of processes will contribute to the 
observed event rate. The goal of the optimisation procedure discussed in the 
next section is to restrict  the phase-space region and to propose the 
corresponding event selection criteria which maximize the contribution of the 
elastic ``point-like'' processes (4). Asymptotically, if the processes  (1-3) 
were totally removed then the luminosity measurement based on the corresponding
event sample could be as precise as the luminosity measurement at the $e^+e^-$ 
colliders using the Bhabha process \cite{Bhabha}. Note that if uncertainty due 
to contribution of processes (1-3) could be reduced below $\sim$0.5\% of the 
total rate, the precision of the matrix element calculation would need to be 
improved (e.g. by creating a new event generator using e.g. the helicity 
amplitude formalism developed by P. Kessler \cite{Kessler} and including the 
electromagnetic radiative corrections). 
 
\subsection{Modeling of the background processes}
 
The Pythia event generator \cite{pythia} was used to develop the methods of the
background suppression strategy. The processes which could give rise to  the 
lepton pair signatures are the soft hadronic processes. They have been studied 
at lower collision energies. Since their extrapolation to the LHC energy domain
is highly uncertain, therefore any  precision method of the LHC luminosity 
measurement must not depend upon their modeling. The Pythia generator is used 
in our studies  merely to define the background subtraction strategy and for 
an initial estimate of the performance requirements for the trigger and data 
selection methods. The background subtraction strategy, discussed in details in
the subsequent paper \cite{paper2},  will  be independent of the Monte Carlo 
modeling of hadronic background processes. The background rates will be 
determined using the dedicated monitoring data samples collected during the 
standard LHC runs.
\begin{figure}
\begin{center}
\epsfig{file=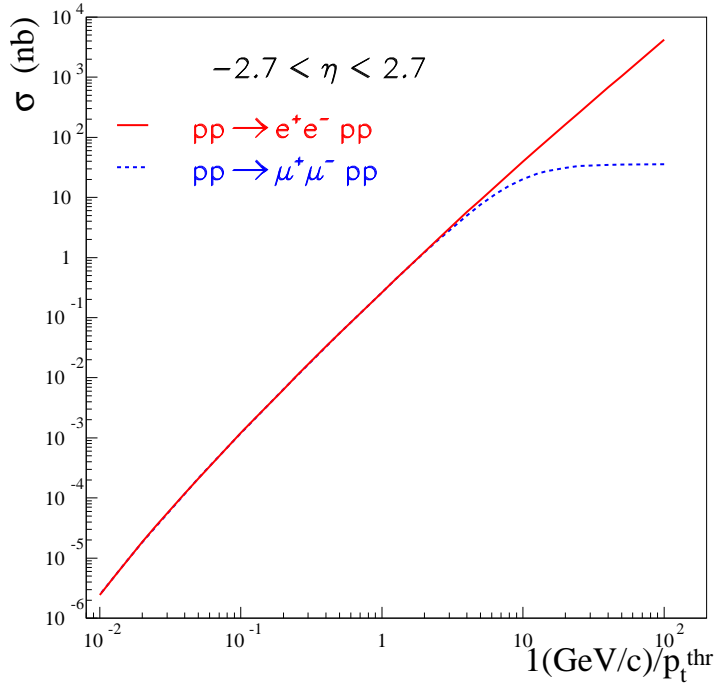,width=10cm}
\end{center}
\caption{\sf The cross section $\sigma$ 
for the lepton pair production in elastic processes  plotted as a 
function  $1{\rm (GeV/c)}/p_{t}^{thr}$ for the sample 
of events  in which the transverse momenta of both leptons
are larger than the transverse momentum threshold,  $p_t^{thr}$. 
The solid line represents  the $e^+e-$ sample while
the $\mu^+\mu^-$ sample  is represented  by the dotted line.}
\label{statistics}
\end{figure}

\section{Optimisation of the measurement region }

\subsection{Statistical precision}

In Fig. \ref{statistics} we show the cross section for the elastic production 
of the muon and electron pairs in the pseudorapidity region 
$-2.7 < \eta < 2.7$, as a function of the inverse of the lepton transverse 
momentum detection threshold $1{\rm (GeV/c)}/p_{t}^{thr}$. In the kinematic 
region studied in this paper, $p_{t}^{thr} \geq m_{\mu}$, the rates of the muon
and electron pairs become identical. As a consequence  the optimisation of the 
kinematic domain presented below is lepton flavor invariant.

The cross section strongly depends on the lepton transverse momentum threshold.
In order to achieve the statistical accuracy of 1\% on daily basis, for the 
average machine luminosity of ${\cal L} = 10^{33}~cm^{-2}s^{-1}$, events with 
the lepton transverse momenta down to to the value of  $\sim$500 MeV/c must be 
efficiently selected. Increasing the lepton $p_t$-detection threshold by one 
order of magnitude reduces the event rate  by three orders of magnitude. In 
such a restricted region 1\% statistical precision could be reached only for a 
one-year-integrated luminosity.

\subsection{Event samples}

In this section  we compare the precision of the theoretical control of the 
lepton pair 
production rates for the two $p^l_t$ detection thresholds: $p_t^{thr1}= 0.2$ 
GeV/c and $p_t^{thr2}= 6$ GeV/c. The corresponding samples of the LPAIR 
generated events will be referred to as the {\bf ``central-track-trigger'' 
(CTT)} and the {\bf ``high-pt-trigger'' (HPT)} samples.

The HPT sample can be collected at LHC by each of the general purpose 
detectors in the muon channel assuming an optimal performance of the trigger 
and the on-line event selection systems. The $p_t^l$ cut is chosen to be the 
same as in 
the analysis presented in \cite{Shamov}. On the other hand, the CTT event 
sample could be collected only if an upgrade of the existing detectors
and their triggers  is made.
  
\begin{figure}[ht]
\begin{center}
\setlength{\unitlength}{1mm}
\begin{picture}(160,80)
\epsfig{file=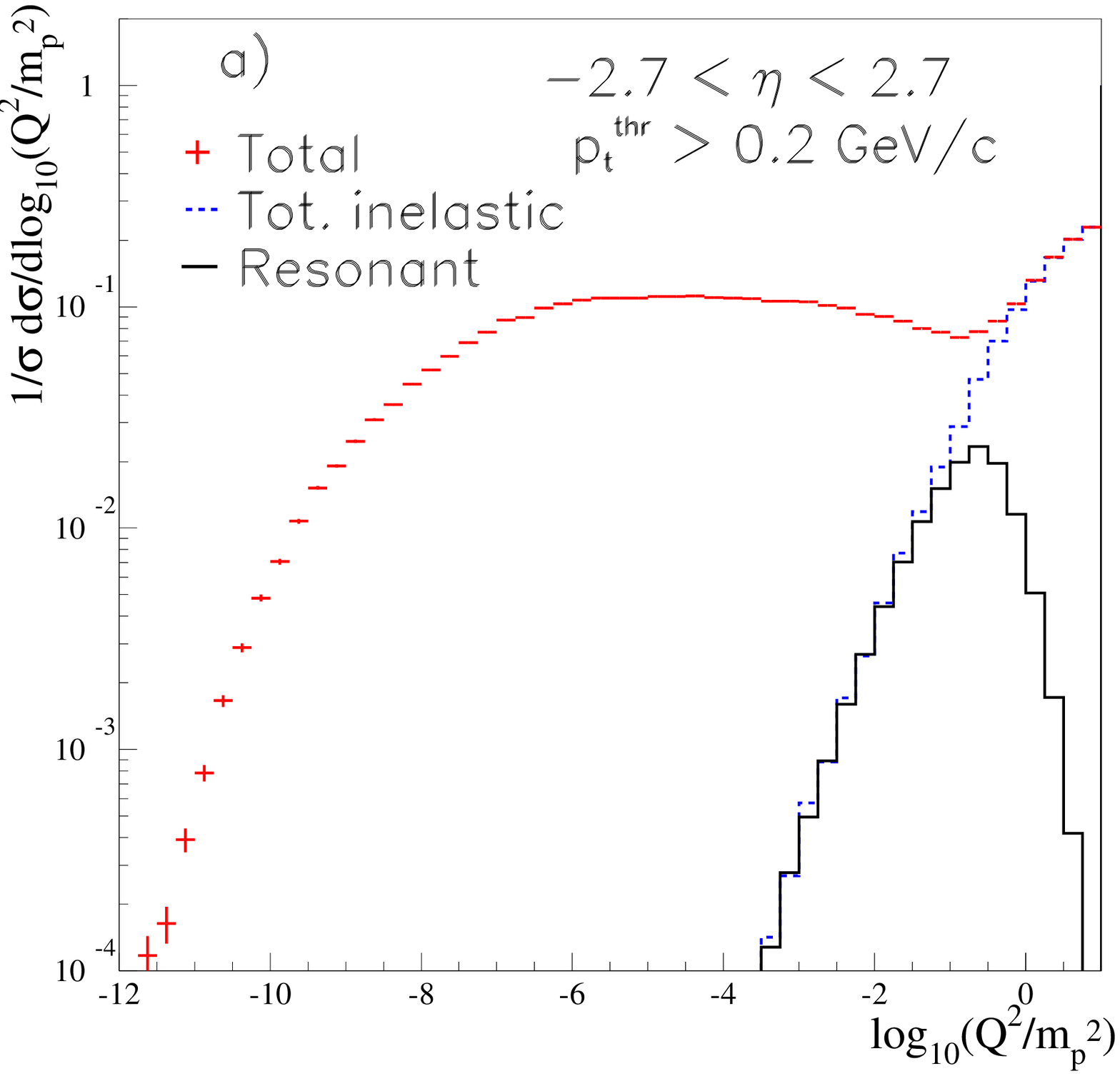, width=75mm}
\epsfig{file=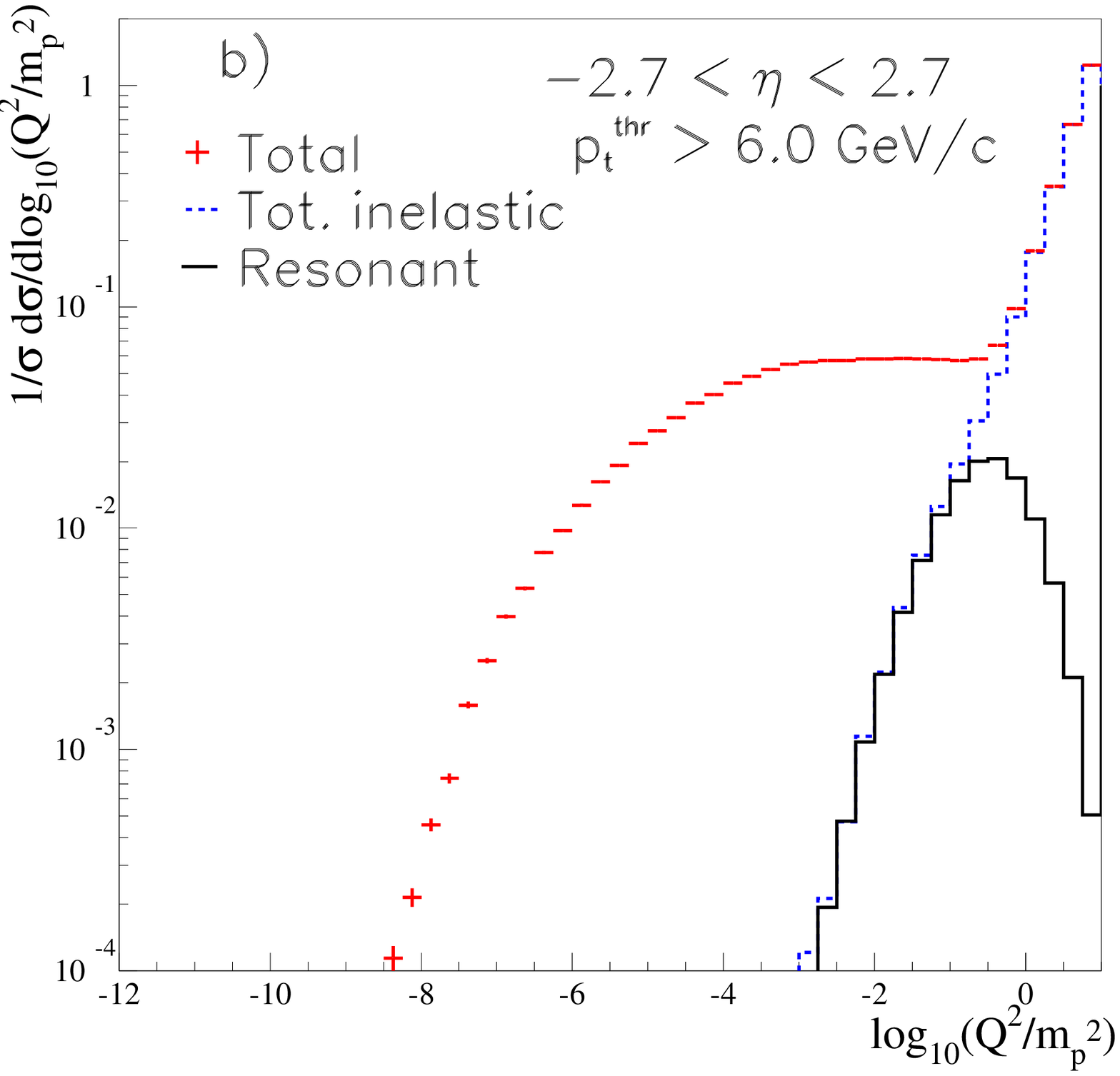, width=75mm}
\end{picture}
\end{center}

\caption{\sf The probability density of the $log(Q^2/m_p^2)$ variable, where  
$Q^2$ are the four-momentum transfer in the proton vertices and $m_p$ is the 
proton mass. The sum of all contributions is represented by crosses. The sum of
all inelastic contributions is shown by line dotted line, while the 
contribution of the low-mass resonant excitations of the proton  are 
represented by the solid line, The distributions are shown separately
for the CTT (a) and for the HPT (b) samples.
Both leptons are produced in the pseudorapidity region  $-2.7 < \eta  <  2.7 $}
\label{q2spectra}
\end{figure}

In order to asses the modeling precision of the absolute rates in the 
selected above kinematic regions we present in Figs. \ref{q2spectra}a,b the 
probability density  
distribution of the $log(Q^2/M_p^2)$ variable, for the CTT and HPT samples. 
The elastic process, as expected, populates the  low $Q^2$ region  while the 
inelastic processes  (including:  the resonance, the photoproduction and the 
deep inelastic
scattering contributions) dominate at large $Q^2$. For the CTT sample (Fig. 
\ref{q2spectra}a) the ratio of the elastic to the  inelastic contribution is 
significantly larger than that for the HPT sample (Fig. \ref{q2spectra}b). 

This is a direct consequence of the fact that the elastic plateau, reflecting 
in the logarithmic scale  the $dQ^2/Q^2$-shape of the 
equivalent photon flux and the strongly peaking invariant mass spectrum of the 
lepton pair, is extended to  lower $Q^2$ values for the CTT sample. This effect
is driven  by the $p_t^{thr}$ cut which determines  the low energy cut-off of 
the spectrum of photons and consequently lowers  the minimal $Q^2$ 
value for the CTT sample.

\subsection{Reduction of modeling  uncertainties}
  
The goal of the kinematic domain optimisation, presented below,  is to maximize
the precision of the theoretical control of the rates while preserving high 
statistical precision of the day-by-day luminosity. To achieve this goal the 
contribution of the processes which are sensitive to the internal proton 
structure  must be drastically reduced. 

An efficient kinematic variable which was used to control the contribution of
the inelastic processes to large angle photon radiation at HERA 
\cite{H1RADIATIVE} was the photon-electron acoplanarity. For LHC, where the 
lepton pair is playing the role of the peripherality tagger  the corresponding 
variable is the lepton pair acoplanarity $\delta\phi$. It  is defined 
as $\delta\phi = \pi - min(2 \pi - |\phi_{1} - \phi_{2}|,|\phi_{1}-\phi_{2}|)$ 
where $\phi_{1}$ and $\phi_{2}$ are the azimuthal angles of produced leptons. 
The acoplanarity variable is correlated with the transverse momentum of the 
lepton pair $p_t^{pair}$. Since the latter variable is, in turn, strongly 
correlated with the impact parameter of the colliding protons, the small 
acoplanarity pairs originate predominantly from the large impact parameter 
collisions. This region is hardly accessible to the strong interaction 
processes because of their restricted interaction range. For the small 
acoplanarity pairs the inelastic excitation of protons in electromagnetic 
processes is also suppressed due  to the QED gauge-invariance which forces the 
disappearance of the photon coupling to inelastic excitations of protons in the
limit of decreasing photon transverse momentum. Because of the above effects 
the production rate of coplanar pairs in processes which resolve the proton 
electric and colour charge structure is drastically reduced\footnote{Note,
that, in general, the large  $Q^2$ value is not necessarily correlated with 
large transverse momentum  of the lepton pair. This correlation becomes strong 
only for the processes in which the energy transfer from the proton to the 
virtual photon is significantly smaller than the initial proton momentum.}.

\begin{figure}[!ht]
\begin{center}
\setlength{\unitlength}{1mm}
\begin{picture}(170,80)
\epsfig{file=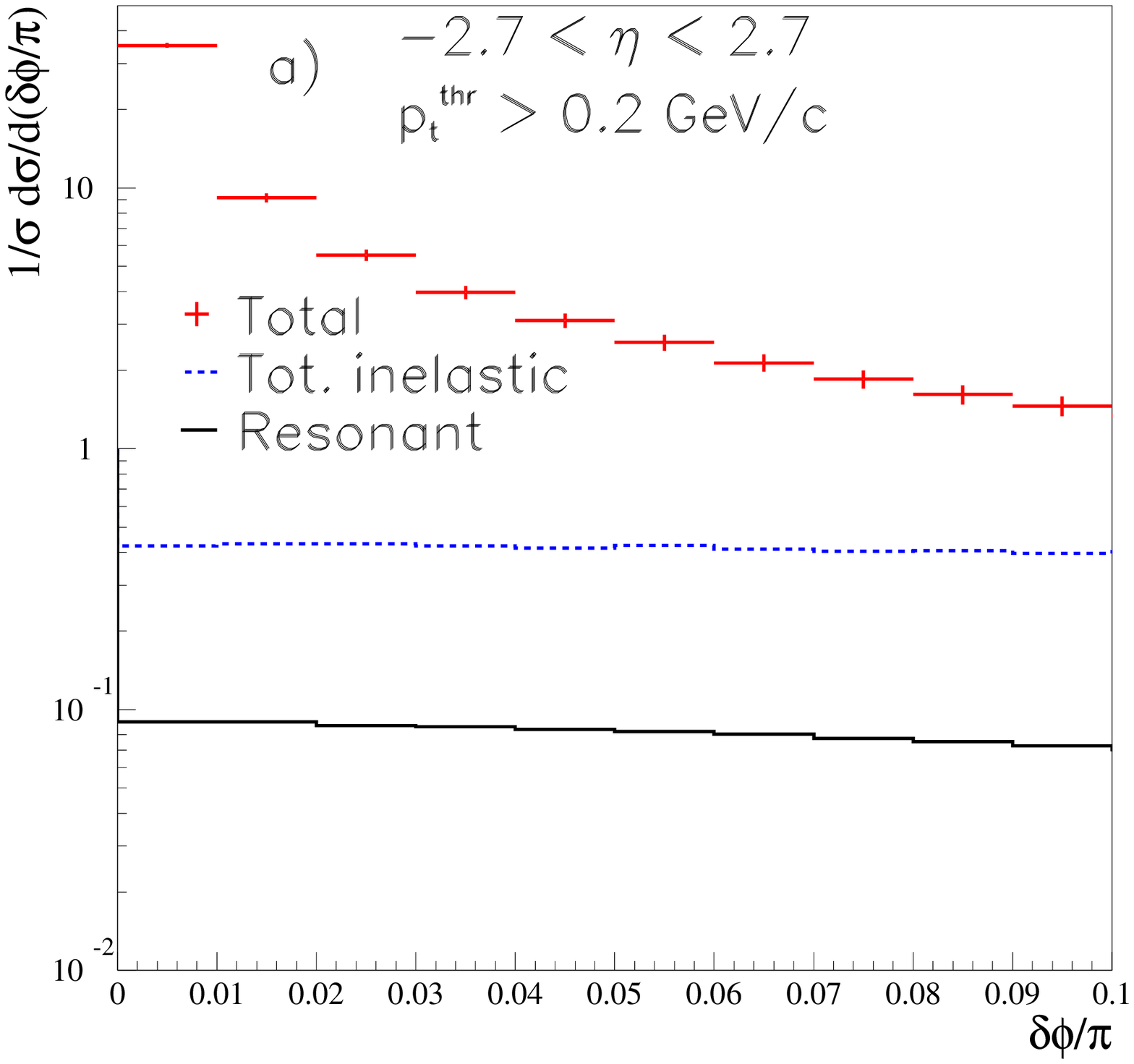, width=75mm}
\epsfig{file=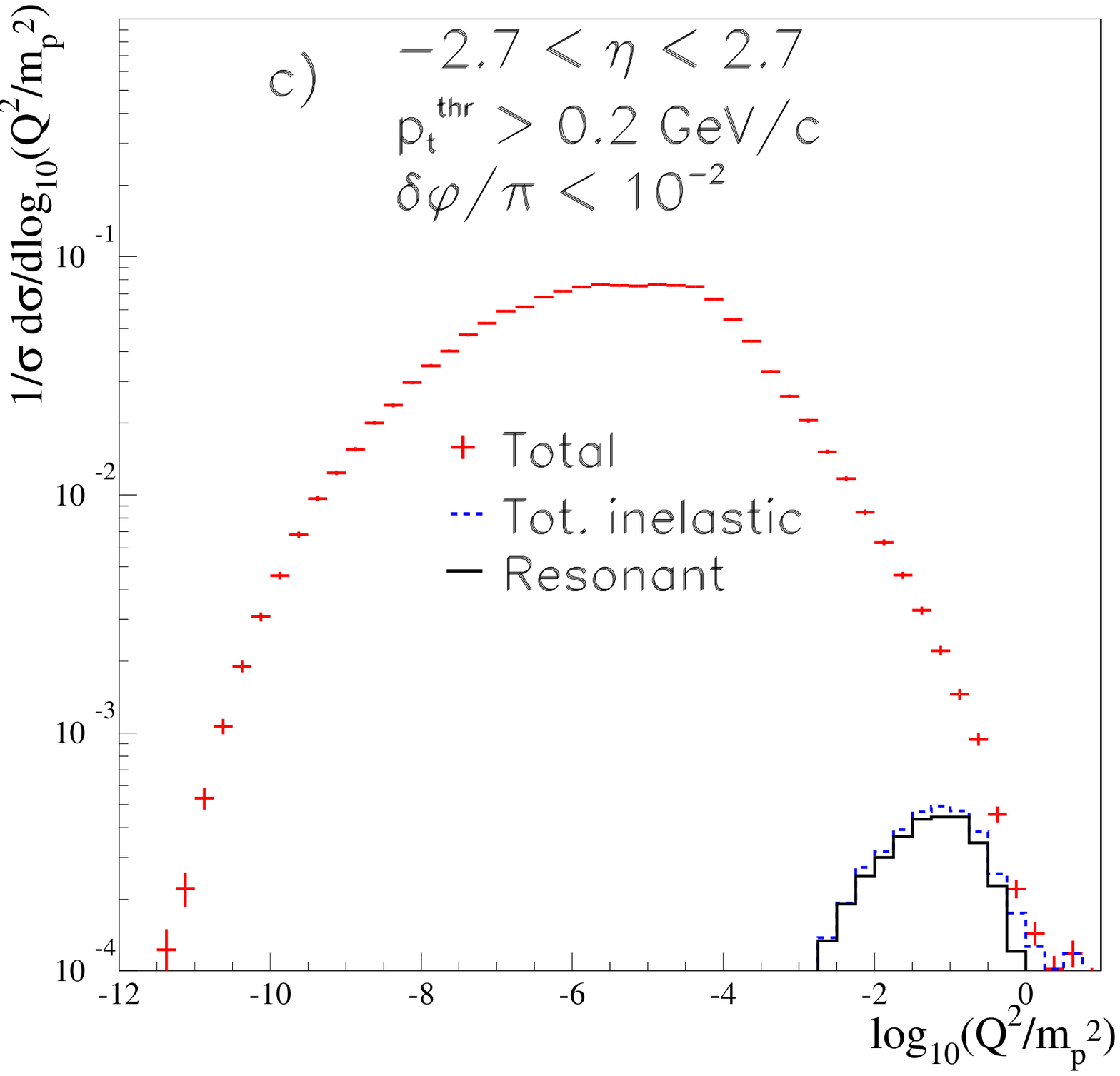, width=75mm}
\end{picture}
\begin{picture}(170,80)
\epsfig{file=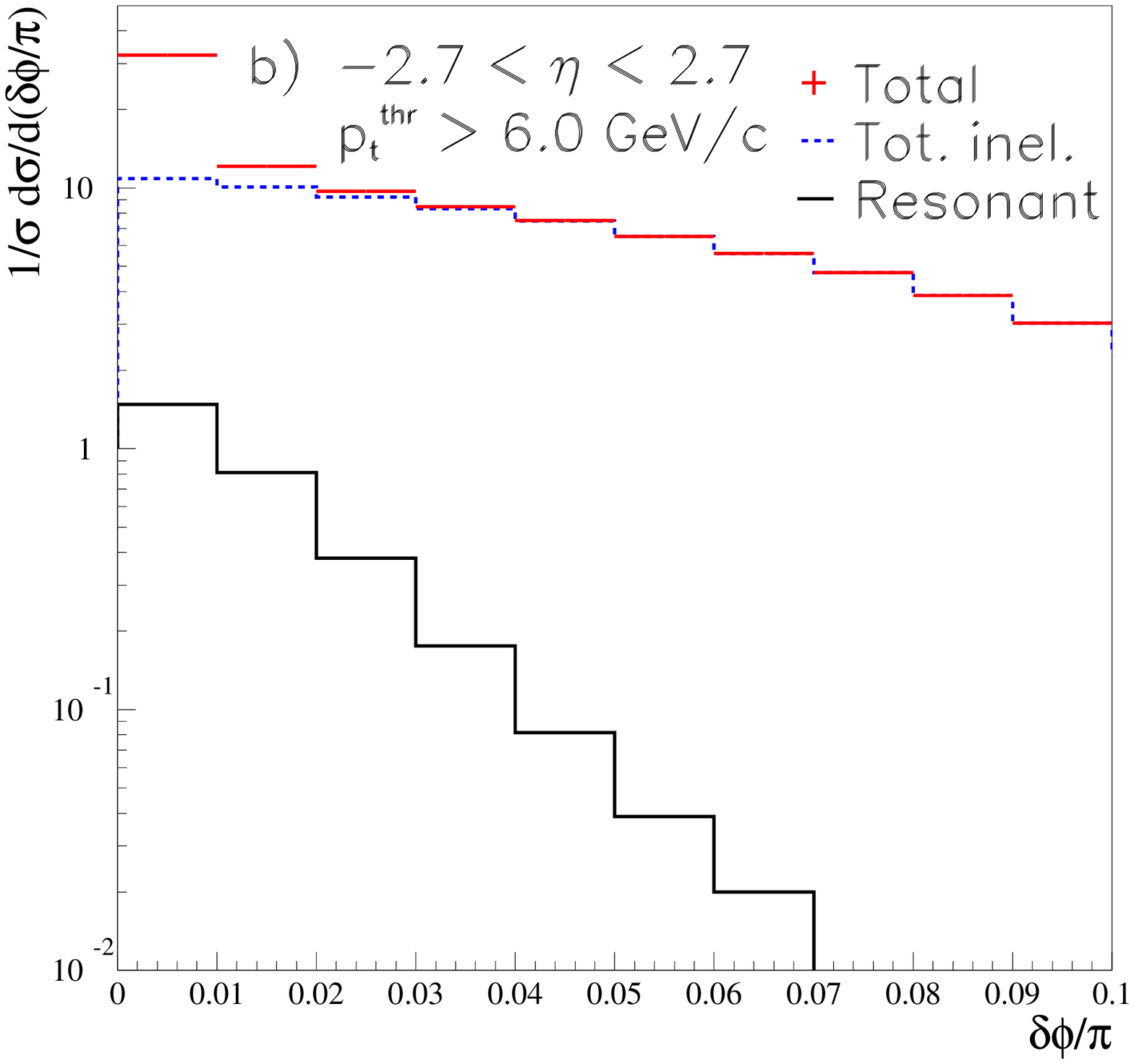, width=75mm}
\epsfig{file=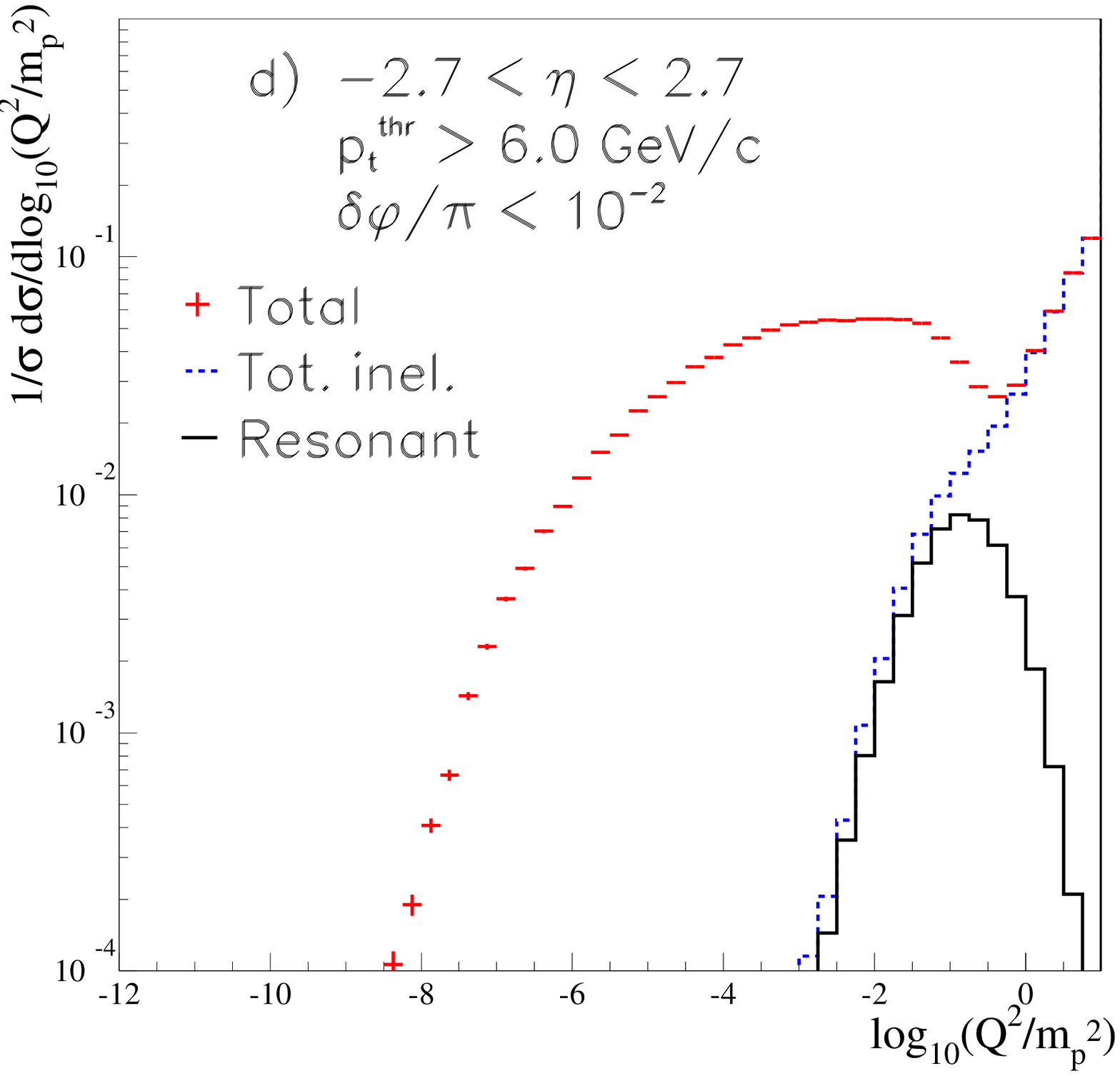, width=75mm}
\end{picture}
\end{center}
\caption{\sf The acoplanarity distribution for lepton pairs produced in the 
pseudorapidity range: $-2.7<\eta <2.7$  for the CTT (a) and the HPT (b) 
samples. The distribution of the logarithm of the four-momentum transfer 
squared, $log(Q^2/m_p^2)$, for events satisfying the condition  
$\delta\phi/\pi\leq  0.01$  for the CTT (c) and for the HPT (d) samples. The 
sum of all contributions is represented by crosses, the sum if inelastic 
contributions is shown by dotted line, and the resonance contribution by solid 
line.} 
\label{accocut}
\end{figure}

In Fig. \ref{accocut} we present the acoplanarity distributions for the CTT
(5a) and for the HPT (5b) samples and the corresponding $Q^2$ distribution for 
the leptons pairs satisfying the conditions $\delta\phi/\pi\le 0.01$,  (5c) and 
(5d). The acoplanarity distribution for the CTT sample is strongly peaked at
$\delta\phi/\pi \simeq 0$ for elastic events and flat for inelastic ones. 
As expected,  the 
acoplanarity cut removes efficiently the inelastic events. 
The acoplanarity distribution for the HPT sample is peaked
at low acoplanarity both for the elastic and  inelastic events. Even if 
a large fraction of inelastic events is removed by the $\delta\phi/\pi\le 0.01$
cut, as shown in Fig. \ref{accocut}d, their remaining contribution, comparable 
to the elastic one, remains. 

In order to achieve better than 1\% precision of the theoretical control of the
event rates we require the contribution of all inelastic processes to be 
reduced  below the level of 20\% of the elastic  event rate, and the 
contribution of those of elastic processes which are sensitive to the proton 
elastic form factors, to be reduced below the level of  50\% of the 
hypothetical ``point-like'' event rate. The ``point-like'' rate  is defined 
here as the pair production rate  for a beam of spin 1/2, point-like particles 
having the mass, the charge, and the anomalous magnetic moment of the proton. 
It can be calculated with precision limited only by the theoretical 
precision of the matrix element $\gamma^*\gamma^* \rightarrow l^+l^-$ and its 
higher order corrections. 

\begin{figure}[!ht]
\begin{center}
\setlength{\unitlength}{1mm}
\begin{picture}(170,80)
\epsfig{file=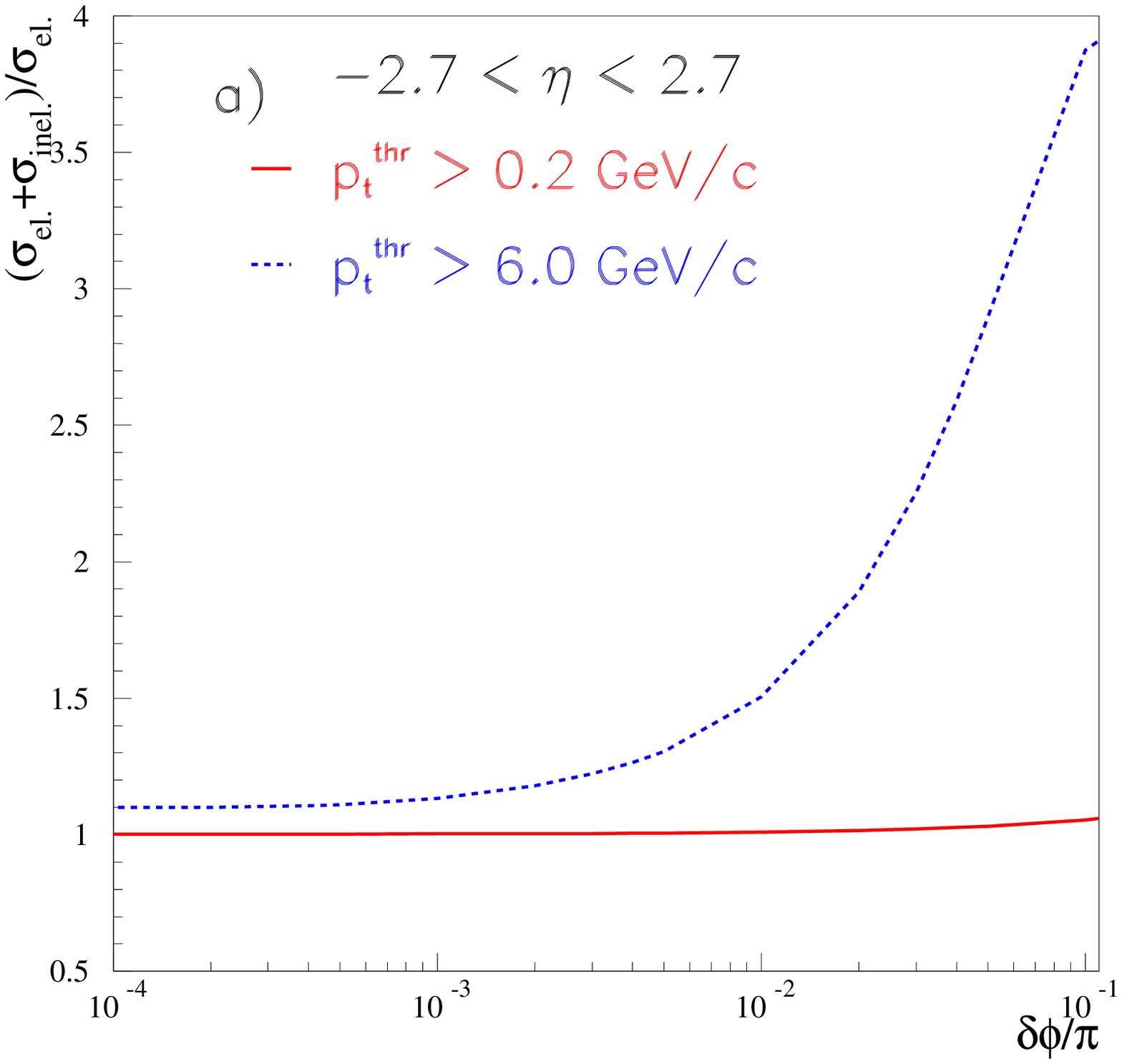, width=75mm}
\epsfig{file=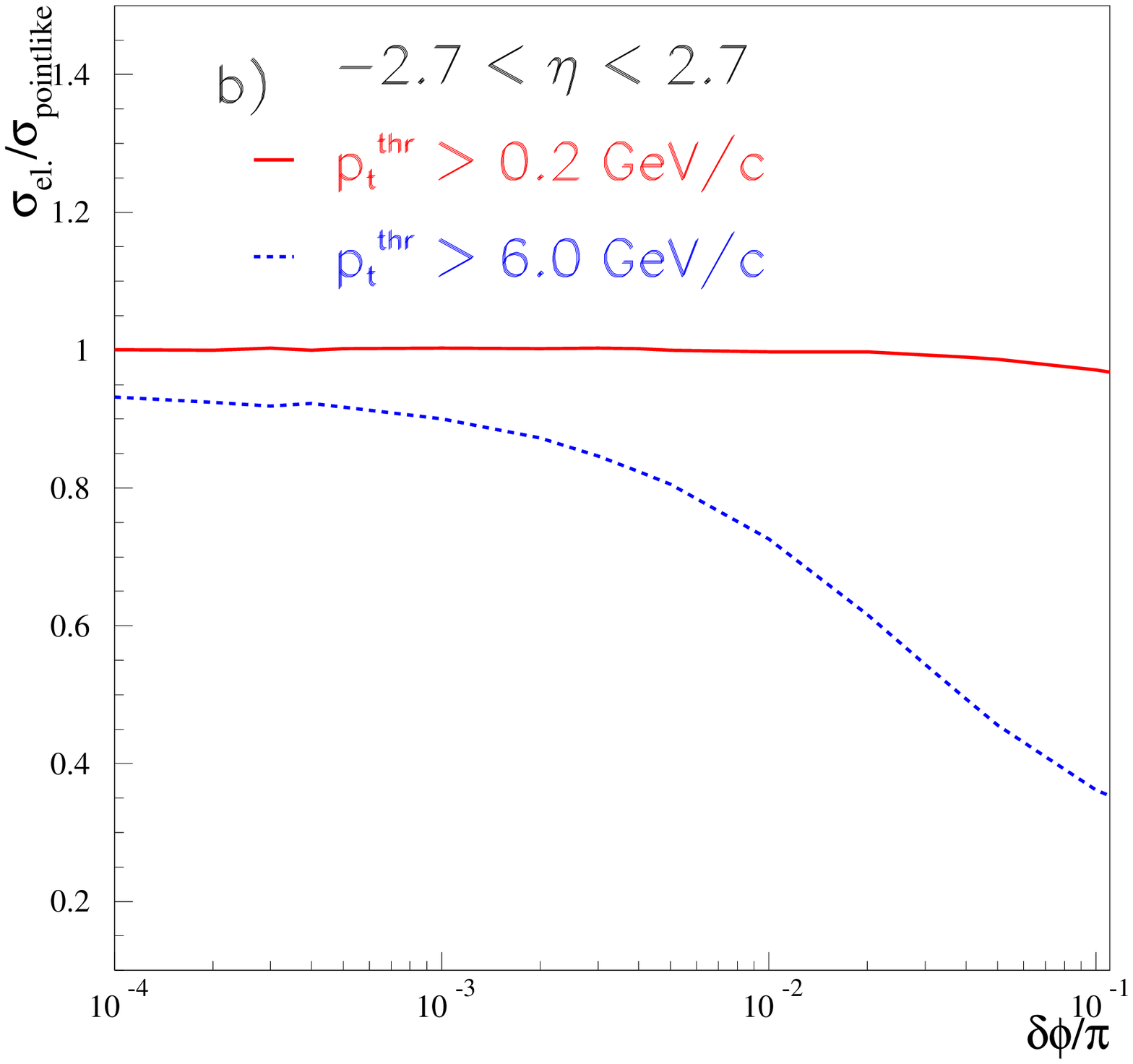, width=75mm}
\end{picture}
\end{center}
\caption{\sf 
The ratio of the total cross section for the lepton pair production to the 
elastic one for the CTT (solid line) and the HPT (dotted line) samples as a 
function of the upper limit on the lepton pair acoplanarity (a).
The ratio of the elastic cross section for the lepton pair production to 
the ``point-like'' one for the CTT (solid line) and the HPT (dotted line) 
samples as a function of the upper limit on the lepton pair acoplanarity (b).
}
\label{elastic}
\end{figure}

In Fig. \ref{elastic}a we show the ratios of the total pair-production cross 
section to the elastic cross section for the CTT (solid line) and for the HPT 
(dotted line) samples as a  function of the upper limit of the accepted 
acoplanarity of a pair. For the CTT sample, already for a loose   
$\delta\phi /\pi\leq 0.1$ cut, the contribution of the inelastic processes  is 
ten times smaller than  of the elastic one. For the HPT sample an equivalent 
acoplanarity cut which reduces the contribution of inelastic processes below 
20\% of the elastic one is $\delta\phi /\pi\leq 0.002$. Note that the 
asymptotic behaviour of 
the ratios at small acoplanarities reflects the difference in the  
lepton transverse momentum cuts for the CTT and HPT sample. For the same 
acoplanarity larger transverse momenta of the lepton pairs are accepted  
for the latter sample.

In Fig. \ref{elastic}b we show the ratios of the elastic section to the 
``point-like'' cross section for the CTT (solid line) and for the HPT (dotted 
line)
samples as a function of the upper limit of the acoplanarity of a pair. 
For the CTT sample, the contribution to the elastic 
cross section sensitive to the proton form factors stays below 20\% of the 
``point-like''  cross section regardless  of the acoplanarity cut. For the HPT 
sample, the contribution to the elastic cross section,  sensitive to the proton
form factors,  is reduced below the required value of 50\% only if the initial 
event sample is restricted  to a  subsample of events satisfying the condition:
$\delta\phi /\pi  \leq 0.03$. Note  that in the case of the HPT sample the 
decrease of the 
elastic contribution at high $Q^2$ values is driven by the proton elastic form 
factors while for the CTT sample by the $p_t^{thr} = 0.2$ GeV/c cut. The latter
cut effectively reduces the contribution of the elastic processes in which the
proton dipole structure is resolved. It is responsible for faster convergence,
of the CTT sample ratio to the value of one with decreasing acoplanarity.

The statistical accuracy of predicted event rate  decreases if the  
acoplanarity range is restricted. This effect is very weak down to the 
acoplanarities  of $\delta\phi/\pi\leq 0.1 $ but becomes significant for 
$\delta\phi/\pi\leq 0.002$ leading to a decrease of the rate of
the HPT events  by a factor of $\sim$4.

\subsection{Constraints} 

We have shown in the previous chapter that for both the CTT and the HPT samples
the contributions of processes sensitive to the internal structure of the beam 
particles can be reduced to a level which assures better than 1\% precision of 
the theoretical control of their rate. However, the luminosity measurement 
based on the HPT sample is vulnerable to several effects which would, very 
likely, reduce its final precision. The more restrictive acoplanarity cut would
require a significantly better understanding of the resolution tails and biases
of the reconstructed azimuthal angles of produced 
lepton pairs for the HPT sample with respect for the CTT sample. The 
corresponding 
reduction of the event rate would extend to, at least, one year the integration
time of rates in order to achieve a statistical precision comparable to the 
theoretical one at ${\cal L} = 10 ^{33}~ cm^{-2} s^{-1}$.
The integration time would  very likely  have to be extended even further if 
the trigger efficiency for the leptons produced close to the $p_t=6 $ GeV/c 
detection threshold would turn out to  be low, or if the 
first-level trigger rate for the hadronic 
events satisfying the lepton pair selection criteria would be higher than 
$\sim 10$ kHz. The latter condition excludes a precise luminosity measurement 
using the HPT electron pairs \cite{ATLAS_TDR} and puts a stringent constraint 
on the maximal acceptable rate of muons coming from the pion and kaon decays. 
The rate gain at the high luminosity ${\cal L} = 10 ^{34} cm^{-2} s^{-1}$ may
lead to a decrease rather than to an increase of the statistical 
precision because the trigger would have to couple with faster increase of 
the background rate with respect to signal one\footnote{The significance of 
the above problems could be diminished if the first level  trigger of the 
LHC experiments would be upgraded from the object-multiplicity scheme to the 
object topology scheme. Such an upgrade is, to our best knowledge, presently
not planned.}.

For the luminosity measurement using the CTT sample there is no ``brick-wall'' 
limit of the reachable theoretical accuracy. The contribution of the inelastic 
and the elastic, form factor dependent,  processes can be 
reduced further without reducing  the signal event rate. A large cross section
for this sample, allowing to collect $10^6$ events over the period of one year 
of running at ${\cal L} = 10 ^{33}~cm^{-2}s^{-1}$ , would assure not only short
luminosity sampling intervals but,  more importantly, a precise control of the 
measurement systematic errors. The increasing systematic precision could be  
matched by increasing precision of the matrix element calculation for 
``point-like''  processes which is  controlled entirely by the cut-off of its 
perturbative expansion. As the example of the LEP experiments has shown,  
a significantly better precision of the luminosity measurements was achieved 
than had initially been anticipated. This is the main reason that we shall 
concentrate in this and subsequent papers only on the luminosity measurement 
method based on the CTT sample being, in our view,  the method which could  
ultimately provide the highest precision for the absolute normalisation of the 
LHC measurements.

At present none of the LHC experiments has the capacity to trigger the CTT  
events and to filter them out with required efficiency. In order to select 
these events an upgrade of the existing detectors is indispensable. Any upgrade
proposal will have to face not only the challenges specific to the LHC machine 
environment but also the challenges of its incorporation within the environment
of the LHC detectors, which were optimised for precise measurements of the 
large  transverse momentum particles. These experimental challenges facing any 
upgrade concept  are discussed below.   
 
\section{ The experimental challenges} 

\subsection{Rejection of strong interaction background}

\begin{figure}[!ht]
\begin{center}
\setlength{\unitlength}{1mm}
\begin{picture}(170,80)
\epsfig{file=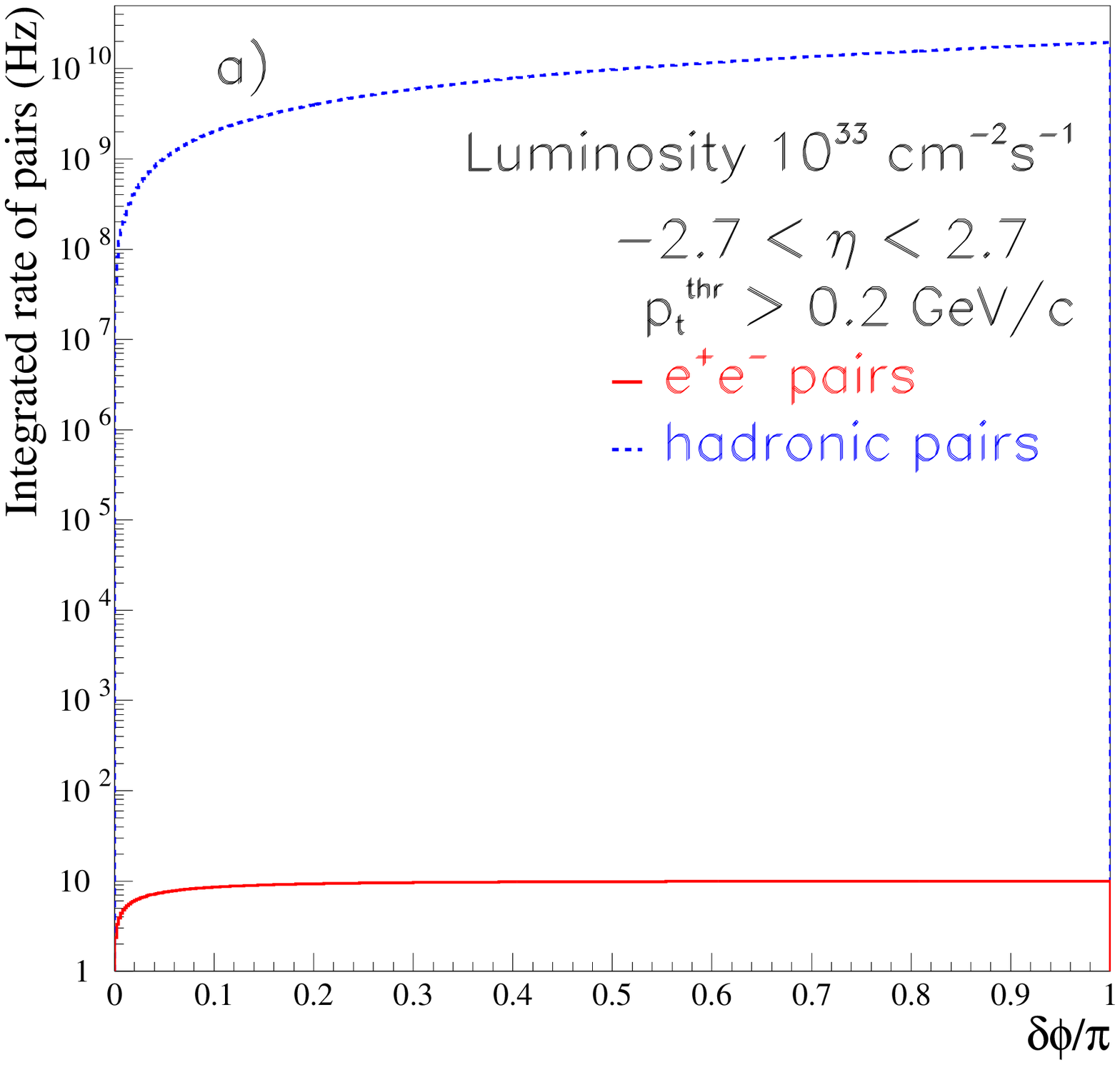, width=75mm}
\epsfig{file=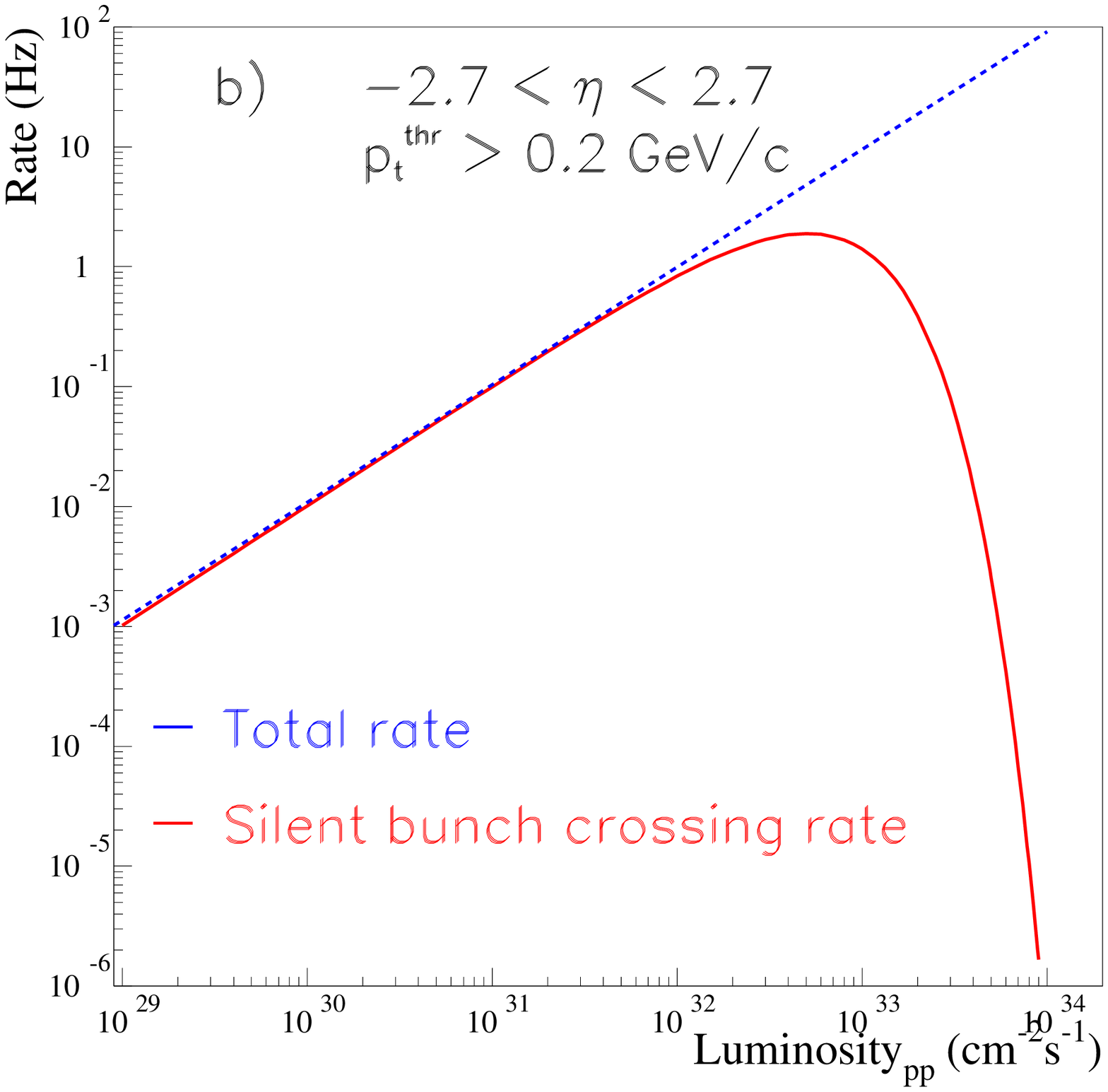, width=75mm}
\end{picture}
\end{center}
\caption{\sf 
The integrated rate of the charged particle pairs as a function of the 
acoplanarity angle for the CTT sample (solid line) and the background (Pythia) 
sample (dotted line) for the luminosity ${\cal L} = 10^{33}$ cm$^{-2}s^{-1}$ 
(a). The dependence of the CTT event rate on the proton--proton luminosity 
(b). The  ``silent bunch crossing'' rate (see text for details) is represented 
by the solid line  and the total rate is represented by the dotted line.} 
\label{rate}
\end{figure}

In Fig. \ref{rate}a we show the rate of the opposite charge particle pairs in 
the pseudorapidity region $-2.7 < \eta < 2.7$ as a function of the upper limit 
of the allowed pair acoplanarity. The rate for the CTT sample (solid line) is 
compared to the corresponding one for the multi-particle production processes 
modeled with Pythia (dotted line). All stable, unlike-charge particle pairs 
produced in the pseudorapidity range $-2.7 < \eta < 2.7$ with transverse 
momenta larger than 0.2 GeV/c contribute to these rates. This plot illustrates 
a basic difficulty in selecting the peripheral electromagnetic collisions. 
For the luminosity of ${\cal L} = 10 ^{33}~cm^{-2}s^{-1}$, 
the predicted,  integrated rate of the signal pairs  is at the Hz level  
while that of the background pairs  reaches the GHz level. 
The challenge is twofold and concerns both the overall rate of background
pairs and their efficient rejection.
First of all, the design of the 
dedicated trigger, the on- and off-line data selection method. must assure   
the overall  background pair rejection of nine orders of magnitude. However, 
the most important challenge is to achieve a sufficiently large  
suppression of the background rates already 
at the very early stage of the data selection process - preferentially 
at the first trigger level operating at large frequency. 

\subsection{Event pile-up}

The signal and the background rates shown in  Fig. \ref{rate}a are calculated  
for continuous beams of colliding particles. In  reality,  the  LHC beams 
are bunched. For the nominal LHC bunch collision 
frequency  and the nominal luminosity several interactions  
occur simultaneously within a bunch crossing. 

In Fig. \ref{rate}b we show the luminosity dependence of the total integrated 
rate of the CTT events and the rate of those of CTT events which are produced 
in the ``silent  bunch crossings'' defined as the bunch crossings with no 
pile-up hadronic collisions. We assumed $\sigma_{pp}^{tot}$ = 79 mbarn and that
the total machine luminosity  will be distributed uniformly to all the LHC 
bunch-crossings occurring with the frequency of 40 MHz.
This plot demonstrates that already for the luminosities lager   than   
${\cal L} \geq 2 \times 10^{31}~cm^{-2}s^{-1}$ the lepton pair
signal events have a non-negligible probability to overlap
with minimum bias interactions. 

The challenge specific to the  bunched beam is thus to develop the trigger and 
the on-line selection methods for the CTT lepton pairs  which are  flexible 
enough to work in  a luminosity-dependent environment of synchronous,  
minimum bias beam collisions.

\subsection{Coplanar lepton pairs in magnetic field }
\begin{figure}[ht]
\begin{center}
\setlength{\unitlength}{1mm}
\begin{picture}(170,80)
\epsfig{file=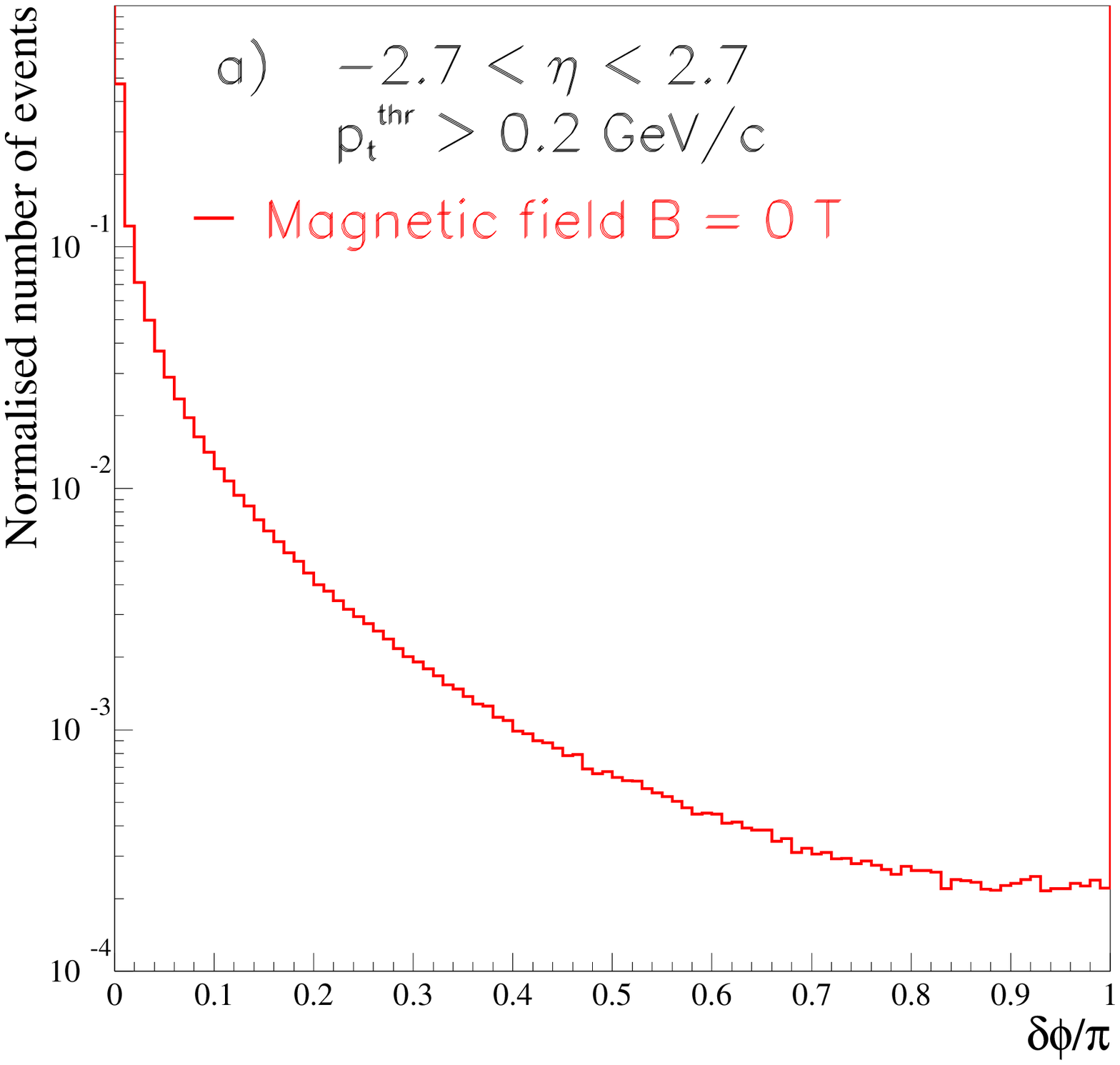,width=80mm}
\epsfig{file=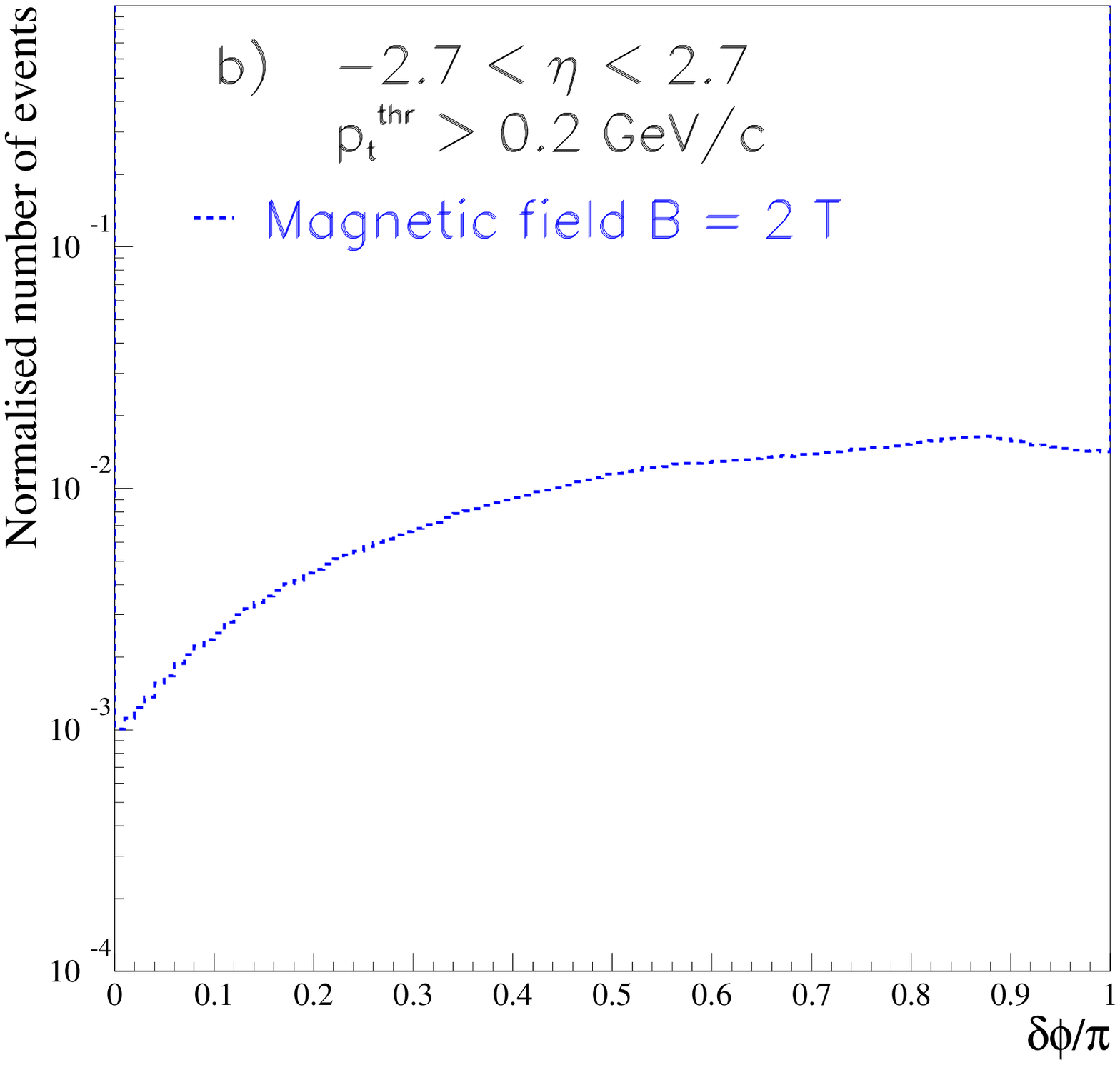,width=80mm}
\end{picture}
\end{center}
\caption{\sf The acoplanarity distribution of the intersection 
of the lepton trajectories with the  
$z=330$ cm plane perpendicular to the colliding beam axis 
 for the CTT event sample: for $B =0$ Tesla (a) , and for $B=2$ Tesla (b).}
\label{field}
\end{figure}

The lepton pairs in the CTT event sample  are coplanar. On the other hand, the 
background pairs coming from strong interaction processes are uniformly 
distributed over the allowed acoplanarity range. This distinction could be used
at the early stage of the event selection process. However, designing an 
efficient trigger for coplanar, small transverse momenta leptons is by no means
straightforward. The standard LHC detector's triggers have been optimized for 
large energy depositions (electron trigger) and high transverse momentum tracks
(muon trigger). Any upgrade project  must provide not only the extension of 
the triggering scheme and  use the signatures of small transverse momentum 
particles but, what is more challenging, it must face the fact that  
these particles  will traverse the region of strong magnetic field.  

To illustrate the influence of the magnetic field on the initially-coplanar 
lepton pairs we show  in Fig. \ref{field} the lepton pair acoplanarity 
distribution on the plane perpendicular to the beam collision axis at the 
distance of  $z=286$ cm from the interaction point for the two values of the 
solenoidal magnetic field of $B =0$ and $B=2$ Tesla. This plot shows that the 
magnetic field essentially  de-correlates the pair's initial acoplanarity. As a
consequence  the lepton's  hit positions cannot be  directly used by the event 
selection algorithm as the indicator of the initial pair acoplanarity. Thus, 
the  challenge for triggering the coplanar pairs is to develop a very fast and 
efficient method of deriving their initial (interaction vertex)  acoplanarities
from the suitable space-time snapshots of the B-field and lepton pair  
kinematic dependent  acoplanarity evolution.

\section{Conclusions and outlook}

In this paper we have studied the lepton pair production process 
$pp \rightarrow l^+l^-  +X$ and its merits for the high precision luminosity 
measurement at the LHC collider. We have selected the kinematic region in which
the rate of the lepton pairs produced in peripheral collisions of the beam 
particles is large enough to achieve a 1\% statistical precision of the 
luminosity measurement on the day-by-day basis. We have demonstrated that 
better than 1\% precision of the theoretical control of the pair rate can be 
reached by a suitable restriction of the phase-space which suppresses the
contribution of the inelastic collisions and those of  elastic collisions in 
which the internal charge-structure of the protons is resolved.

Selecting  lepton-pairs in the above phase space region represents a major 
challenge which, at present, is beyond the reach of the LHC detectors. In the 
forthcoming paper \cite{paper2} we shall discuss the necessary detector and the
trigger system upgrade to meet this challenge. We shall propose the measurement
strategy  which aims at achieving a comparable systematic measurement precision
to the statical and theoretical ones.   

\end{document}